# A Five-Layer Framework for AI Governance: Integrating Regulation, Standards, and Certification

Avinash Agarwal[1] and Manisha J. Nene[2]

[1]Department of Telecommunications, Ministry of Communications, Sanchar Bhawan, New Delhi, India

[2]Defence Institute of Advanced Technology, Ministry of Defence, Girinagar, Pune, India

**Abstract**

**Purpose** - The governance of artificial intelligence (AI) systems requires a structured approach that connects high-level regulatory principles with practical implementation. Existing frameworks lack clarity on how regulations translate into conformity mechanisms, leading to gaps in compliance and enforcement. This paper addresses this critical gap in AI governance.

**Methodology/Approach** - A five-layer AI governance framework is proposed, spanning from broad regulatory mandates to specific standards, assessment methodologies, and certification processes. By narrowing its scope through progressively focused layers, the framework provides a structured pathway to meet technical, regulatory, and ethical requirements. Its applicability is validated through two case studies on AI fairness and AI incident reporting.

**Findings** - The case studies demonstrate the framework's ability to identify gaps in legal mandates, standardization, and implementation. It adapts to both global and region-specific AI governance needs, mapping regulatory mandates with practical applications to improve compliance and risk management.

**Practical Implications** - By offering a clear and actionable roadmap, this work contributes to global AI governance by equipping policymakers, regulators, and industry stakeholders with a model to enhance compliance and risk management.

**Social Implications** - The framework supports the development of policies that build public trust and promote the ethical use of AI for the benefit of society.

**Originality/Value** - This study proposes a five-layer AI governance framework that bridges high-level regulatory mandates and implementation guidelines. Validated through case studies on AI fairness and incident reporting, it identifies gaps such as missing standardized assessment procedures and reporting mechanisms, providing a structured foundation for targeted governance measures.

## 1 Introduction

Artificial Intelligence (AI) is increasingly becoming a cornerstone of technological progress, influencing diverse sectors such as telecom, healthcare, finance, transportation, defense, and public services. As AI systems continue to permeate these critical domains, the need for robust governance mechanisms has never been more urgent. AI governance refers to the set of policies, regulations, standards, and practices designed to ensure that AI technologies are developed, deployed, and utilized ethically, transparently, and responsibly. The desired outcomes of AI governance include ensuring fairness, preventing harm, fostering innovation, and maintaining public trust in AI systems [Dafoe, 2018].

However, the rapidly evolving nature of AI presents significant challenges to traditional governance models. The complexity and opacity of many AI systems, combined with their potential for widespread societal impact, require a governance approach that is both flexible and comprehensive. The European Union's Artificial Intelligence Act (EU AI Act) [European Parliament, 2024] and the National Institute of Standards and Technology AI Risk Management Framework (NIST

---

[1] Corresponding author: avinash.70@gov.in

AI RMF) [NIST, 2023] are notable examples of regulatory efforts towards addressing these challenges. While these frameworks provide crucial guidelines and risk management strategies for AI, they do not provide the detailed procedural guidance necessary for consistent implementation across various sectors and applications [Ronanki et al., 2023].

This gap between high-level regulatory mandates and the specific, actionable guidance needed for practical implementation is a critical issue in AI governance [Georgieva et al., 2022]. Regulatory frameworks like the EU AI Act may mandate fairness assessments for high-risk AI systems but often stop short of specifying the methodologies, benchmarks, or tools required to perform these assessments. This lack of detail can lead to inconsistencies in how AI systems are evaluated, potentially undermining the effectiveness of the regulations and leading to a lack of trust in AI technologies [Ronanki et al., 2023].

This paper proposes a five-layer framework to address these gaps and enhance the effectiveness of AI governance. In this paper, AI governance refers to a structured approach that spans regulatory frameworks, technical standards, assessment methodologies, and certification mechanisms to ensure AI systems are robust, trustworthy, and accountable. Unlike corporate AI governance, which focuses on internal policies and risk management, our approach addresses governance at a broader level, linking laws, regulations, and standardized evaluation processes to create a cohesive oversight structure for AI systems. The proposed framework systematically connects high-level policies with the detailed implementation processes necessary for effective oversight. The strength of the five-layer framework lies in its ability to provide increasingly detailed guidance as one moves down the layers while simultaneously narrowing the focus to specific aspects of AI governance. For example, while the first layer may broadly require that AI systems be fair, the second layer might focus specifically on developing standards for fairness in AI, the third on establishing procedures for assessing fairness, the fourth on creating tools for this assessment, and the fifth on certifying that the system meets the established fairness criteria. This structured approach ensures that AI governance is not only comprehensive but also adaptable to different AI applications and contexts.

The framework's practical applicability and versatility are demonstrated through two case studies: AI fairness and AI incident reporting. These were deliberately selected to reflect two contrasting domains: one that is well-researched and actively regulated and another that has received limited regulatory and academic attention. AI fairness has seen significant global focus from researchers, regulators, and standard-setting bodies, making it a suitable candidate to assess how well the framework maps existing efforts, identifies good practices and evaluates the presence of governance elements in each layer. This case also includes a country-specific perspective, using India to show how global frameworks can be adapted to meet national priorities. In contrast, AI incident reporting is underdeveloped in terms of accountability, regulation, and implementation [Lupo, 2023]. Unlike fairness, safety, or security, areas with relatively mature frameworks, AI incident reporting lacks legal mandates, standardized taxonomies, systematic procedures, and certification mechanisms in most jurisdictions [Turri and Dzombak, 2023, Avinash and Manisha, 2024]. This makes it a valuable test case for illustrating how the framework can reveal governance gaps. Together, these two cases offer a balanced validation. Specific contributions of this study include:

1. Proposes a five-layer AI governance framework that ensures a structured transition from high-level regulations to detailed implementation guidelines, enabling comprehensive policy and compliance integration.

2. Validates the framework through case studies on AI fairness and AI incident reporting, demonstrating its applicability across diverse AI governance challenges.

3. Identifies key gaps in AI governance, including the lack of standardized assessment procedures, reporting mechanisms, and certification processes.

4. Provides a structured foundation for policymakers, regulators, and industry stakeholders to develop targeted governance mechanisms, balancing regulatory oversight with practical implementation.

The paper is structured as follows: Section 2 reviews previous work on AI governance frameworks. Section 3 introduces the proposed five-layer framework. Section 4 presents a case study on AI fairness, including global and regional initiatives. Section 5 examines AI incident reporting as a second case study. Both case studies serve as validation of the framework. Section 6 discusses the framework's applicability to other aspects of AI governance and suggests directions for future work. Finally, Section 7 concludes by summarizing the key contributions and impact of the paper.

## 2 Literature Review

The field of AI governance is characterized by a growing consensus on its importance, yet faces significant challenges in practical implementation. A central theme in recent research is the urgency and necessity of establishing effective AI governance frameworks [Bernd W. Wirtz and Sturm, 2020, Taeihagh, 2021, Zaidan and Ibrahim, 2024, Roberts et al., 2024, Hadzovic et al., 2023]. Studies consistently highlight the rapid advancement of AI and the potential risks it poses, emphasizing the need for proactive measures to guide its development and deployment in a responsible manner [Taeihagh, 2021, Roberts et al., 2024, Hadzovic et al., 2023]. This urgency stems from the recognition that current governance mechanisms are struggling to keep pace with the evolving complexities of AI technology [Taeihagh, 2021, Zaidan and Ibrahim, 2024].

While ethical principles are widely discussed as foundational to AI governance, a critical gap exists in translating these ethical considerations into practical and actionable frameworks [de Almeida et al., 2021, Batool et al., 2025, Prem, 2023]. Systematic reviews of the literature reveal a dominant focus on ethical values like fairness, transparency, and accountability [Batool et al., 2025, Prem, 2023]. However, these reviews also point out that many existing proposals remain too abstract and lack concrete guidance for implementation [de Almeida et al., 2021, Prem, 2023]. This gap between ethical ideals and practical application is a recurring concern, indicating a need to move beyond high-level principles and develop more operational frameworks [de Almeida et al., 2021, Batool et al., 2025].

The challenge of standardization and interoperability emerges as a significant hurdle in AI governance [Zaidan and Ibrahim, 2024, Fritz and Giardini, 2024, Clarke, 2019, Smuha, 2021]. Concerns are raised about the potential for fragmented regulatory landscapes and the difficulty of achieving international coordination [Fritz and Giardini, 2024, Smuha, 2021]. The need for unified regulatory frameworks and standardized approaches is highlighted to mitigate the risks of regulatory divergence and ensure consistent governance across different jurisdictions [Fritz and Giardini, 2024, Clarke, 2019]. This standardization challenge is particularly relevant to the development of assessment and certification mechanisms, as these need to be applicable and recognized across diverse contexts and regulatory regimes.

Accountability and enforcement are also critical themes in AI governance research [Novelli et al., 2024, Outeda, 2024]. The concept of accountability itself is debated, with calls for clearer definitions and operationalizations [Novelli et al., 2024]. Examining specific application areas like autonomous robotic surgery, [O'Sullivan et al., 2019] explores the legal, regulatory, and ethical frameworks required, emphasizing the need to address responsibility across accountability, liability, and culpability. Analyses of regulatory efforts like the EU AI Act point to the importance of establishing mechanisms for accountability and enforcement to ensure that ethical principles and regulatory requirements are effectively translated into practice [Outeda, 2024]. However, the practical implementation of accountability mechanisms, including standardized assessment and certification procedures, remains a key area needing further development.

Recognizing these challenges, many studies emphasize the need to move towards more practical tools, methodologies, and best practices for AI governance [de Almeida et al., 2021, Papagiannidis et al., 2023, Zaidan and Ibrahim, 2024, Mesk´o and Topol, 2023, Prem, 2023]. Calls for realworld case studies, empirical analysis, and the development of concrete instruments for assessment and verification are prevalent [de Almeida et al., 2021, Papagiannidis et al., 2023, Prem, 2023]. Specific recommendations for regulatory oversight in areas like healthcare [Mesk´o and Topol, 2023] and organizational AI governance [Papagiannidis et al., 2023] underscore the growing demand for practical guidance and actionable frameworks. This emphasis on practical implementation aligns directly with this paper's focus on establishing a structured framework for standardized assessment and certification in AI governance.

Examining specific regulatory initiatives, such as the EU AI Act [Outeda, 2024], provides valuable insights but also reveals ongoing challenges. While the EU AI Act represents a significant step towards comprehensive AI regulation, its limitations, particularly in practical implementation and potential for self-governance by industry, are noted [Outeda, 2024]. This analysis suggests that even with ambitious regulatory frameworks, ensuring effective implementation, oversight, and standardized assessment remains a crucial and complex task.

One prominent approach to structuring AI governance involves layered and modular frameworks [Gasser and Almeida, 2017, M¨okander et al., 2024, Mesk´o and Topol, 2023, Agarwal and Agarwal, 2024]. These models aim to organize the multifaceted nature of AI governance by dividing it into distinct layers or modules, such as social, ethical, legal, and technical dimensions [Gasser and Almeida, 2017, Agarwal and Agarwal, 2024]. Layered approaches are proposed for various contexts, including general AI governance [Gasser and Almeida, 2017], auditing large language models [Mökander et al., 2024], and regulating AI in healthcare [Meskó and Topol, 2023]. While offering a structured way to conceptualize AI governance, questions remain about how these layers practically interact and how compliance can be consistently assessed across them. Specifically, the challenge of standardizing assessment procedures and certification mechanisms within these layered frameworks is not always explicitly addressed.

Complementary to layered models are integrated and comprehensive frameworks that seek to provide a holistic approach to AI governance [Bernd W. Wirtz and Sturm, 2020, Choung et al., 2024, Agarwal and Agarwal, 2024]. These frameworks emphasize the interconnectedness of different aspects of governance and aim to encompass a wide range of stakeholders and considerations [Bernd W. Wirtz and Sturm, 2020, Choung et al., 2024]. For instance, integrated frameworks for public administration [Bernd W. Wirtz and Sturm, 2020] and multi-level governance models involving governments, corporations, and citizens [Choung et al., 2024] are proposed. A sevenlayer model for standardizing AI fairness assessment [Agarwal and Agarwal, 2024] also exemplifies this comprehensive approach. Despite their aim for holism, these frameworks often require further elaboration on the specific tools and processes needed for standardized assessment and verification of AI systems.

In conclusion, the existing literature on AI governance frameworks, including ethical principles, structural models, and layered approaches, underscores the urgency and complexity of translating high-level ideals into effective, enforceable practice. While layered models offer useful conceptual organization, they often fall short in guiding implementation and detailing how coordination occurs across layers or how assessment and certification mechanisms can be standardized and operationalized in diverse domains. Moreover, challenges remain around sector-specific implementation, practical accountability, and consistent oversight. Addressing these gaps, this paper proposes a five-layer framework that not only integrates regulatory, standardization, and certification mechanisms but also provides a structured, implementable approach that links high-level regulatory intent with actionable tools for compliance and trust-building in AI systems.

## 3   Our proposed five-layer framework

The proposed five-layer framework for AI governance provides a structured approach to managing the complexities of AI systems. It offers a layered approach that progresses from high-level overarching legal principles to concrete implementation mechanisms to ensure a comprehensive governance structure. Each layer serves a distinct purpose, ensuring that AI governance encompasses both broad regulatory guidelines and practical, actionable procedures.

Laws and regulations form the foundation of AI governance, setting fundamental principles at the highest level. These principles are implemented through standards and assessment procedures, ensuring a structured approach. Assessment tools and metrics provide practical methods for evaluation, while a certification ecosystem verifies compliance and builds trust. This layered structure assigns clear responsibilities and brings together key stakeholders within a unified governance framework. The five layers and their primary ownerships are tabulated in Table 1. Fig. 1 depicts the interrelation between these five layers.

Table 1: The five-layer framework for AI governance (Source: Authors' own work)

| Layer | Title | Primary ownership |
|---|---|---|
| 1 | Laws, regulations, and policies | Governments, Multilateral organizations |
| 2 | Standards | Standards organizations |
| 3 | Standardized assessment procedures | Standards organizations |
| 4 | Standardized assessment tools and metrics | Academia, researchers, industry |
| 5 | Certification ecosystem | Developers, auditors |

## 3.1 Layer-1: Laws, regulations, and policies

The first layer of the framework establishes the foundational legal and regulatory principles that govern AI systems. It includes high-level regulations and laws that set the key requirements for AI governance. The goal is to ensure AI systems comply with fundamental legal standards and uphold ethical considerations. Governments and multilateral organizations, such as the EU and OECD, are primarily responsible for defining these regulations. The aim is to protect public welfare by ensuring that AI systems operate within boundaries that address safety, non-discrimination, privacy, and human rights.

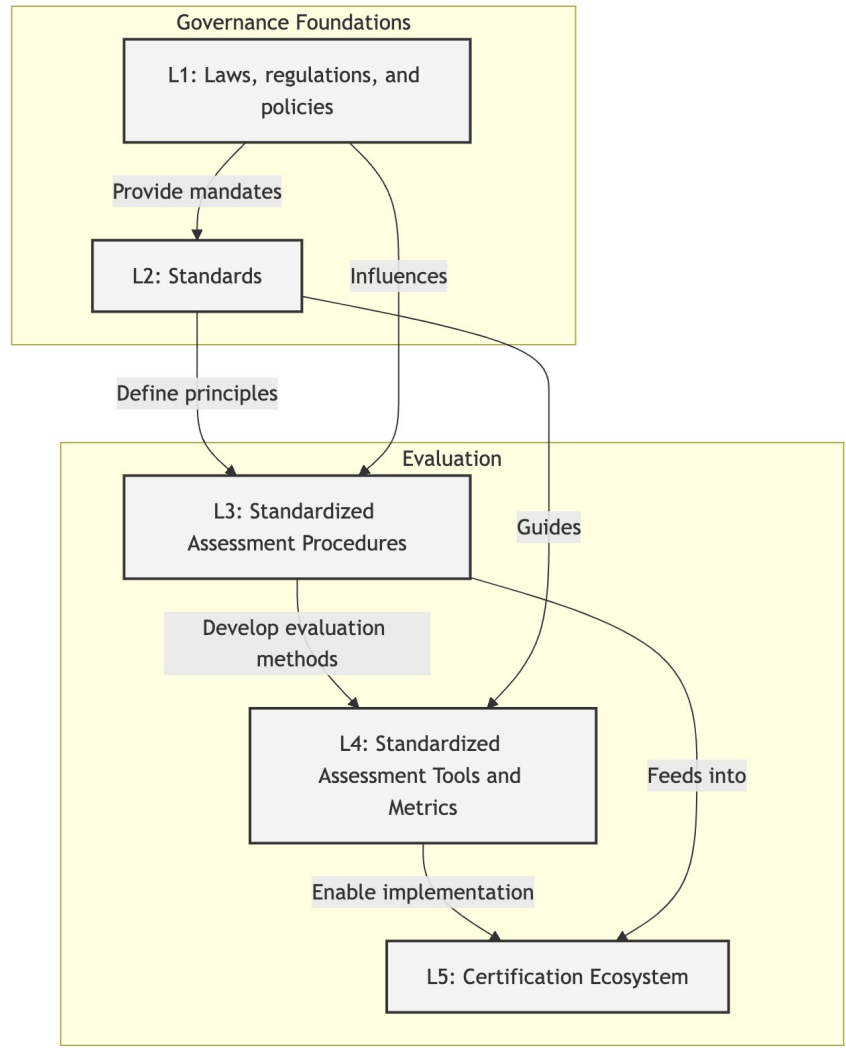

Figure 1: Interrelation between the layers of the proposed framework (Source: Authors' own work)

For example, the EU AI Act classifies AI systems into four risk categories: unacceptable risk, which includes systems banned outright (e.g., social scoring by governments); high risk, covering AI used in critical sectors (e.g., healthcare, transport, employment) with highest regulatory burden and strict compliance obligations; limited risk, where transparency requirements apply (e.g., chatbots); and minimal risk, which faces no additional regulation [European Parliament, 2024]. It also emphasizes ethical principles such as fairness, transparency, and accountability. This illustrates how Layer 1 defines the legal and ethical boundaries within which AI operates, providing guidance for the more detailed standards and procedures addressed in subsequent layers.

## 3.2 Layer-2: Standards

Layer-2 puts the high-level principles from Layer-1 into practice by offering specific standards and guidelines for designing, developing, and deploying AI systems. These standards furnish detailed requirements that AI systems must meet on specialized topics, such as AI fairness. They address both technical and operational aspects like algorithmic fairness, data quality, transparency, and accountability. Standards organizations, such as Institute of Electrical and Electronics Engineers (IEEE), International Organization for Standardization (ISO), International Electrotechnical Commission (IEC), and National Institute of Standards and Technology (NIST), are key in developing these standards. Stakeholders from various sectors, including AI developers, researchers, academia, policymakers, and civil society, should actively contribute to developing the standards to ensure a comprehensive perspective.

Key examples include ISO/IEC 42001, the first AI management system standard for governance [International Organization for Standardization, 2023c]; ISO/IEC 23894, offering specific guidance on managing AI-related risks [International Organization for Standardization, 2023a]; IEEE 7001-2021, part of the broader IEEE P7000 series addressing AI ethics, which specifically establishes measurable and testable levels of transparency for autonomous systems [IEEE Standards Association, 2022a]; the NIST AI Risk Management Framework (AI RMF), providing a voluntary structure for managing AI risks [NIST, 2023]; and the Telecommunication Engineering Centre (TEC) Standard for Fairness Assessment and Rating of Artificial Intelligence Systems [Telecommunication Engineering Centre, 2023]. These standards establish benchmarks for compliance, ensuring that AI systems meet the ethical and operational expectations set out in Layer-1. They also help organizations implement AI in a way that promotes consistency and trust, providing measurable criteria for assessing AI fairness, transparency, and other key attributes.

### 3.3 Layer-3: Standardized assessment procedures

Layer-3 defines the methodologies and procedures for evaluating AI systems based on the standards and guidelines established in Layer-2. The focus is to ensure that evaluations are systematic, consistent, and reproducible. While the standards (Layer-2) provide the requirements, the assessment procedures (Layer-3) define the test methods used to verify compliance with each such requirement. These procedures support the assessment of AI systems against legal, ethical, and technical standards. This layer ensures that assessments are both thorough and replicable.

For example, standardized procedures for fairness assessments or robustness evaluations include predefined methods for measuring AI performance and checking whether it meets established benchmarks. These procedures ensure consistent evaluations across different AI systems and application domains. Several international and national standards provide such structured assessment protocols. ISO/IEC TS 4213:2022 specifies methods to assess machine learning classification performance, focusing on accuracy while ISO/IEC 24029-2:2023 provides a methodology for evaluating neural network robustness against adversarial attacks [International Organization for Standardization, 2022, 2023b]. IEEE 3129-2023 establishes test procedures for assessing the robustness of AI-based image recognition services [IEEE Standards Association, 2023]. Recommendation ITU-T Y.3162 offers procedures to evaluate and grade the intelligence level for network slice management and orchestration in IMT-2020 networks and beyond [International Telecommunication Union, 2024]. India's TEC 57050:2023 outlines a framework for assessing and rating AI fairness, including certification processes [Telecommunication Engineering Centre, 2023].

These standardized assessment procedures offer a common reference for regulators, auditors, and developers, enabling transparent, comparable, and trustworthy evaluations of AI systems across domains.

### 3.4 Layer-4: Standardized assessment tools and metrics

Layer-4 provides the practical tools and metrics required for conducting the assessments defined in Layer-3. These tools and metrics enable measurable and reproducible evaluations of compliance of AI systems with standards and assessment procedures established in the previous layers. Standardizing these tools and metrics is essential for ensuring that assessments in Layer-3 are grounded in reliable and consistent methods so that the assessments have more acceptability. Academia, researchers, and industry stakeholders play a significant role in developing and refining these tools and metrics.

The assessment procedures to check for requirements may need programming or writing software code. Not every organization has the expertise or resources for this purpose. The AI assessment tools are generally pre-built libraries of code, often written in Python, and provide standardized functions for testing specific requirements. Many such toolkits are available as open-source. For example, Adversarial Robustness Toolbox (ART) [Nicolae et al., 2018] is popular for evaluating AI robustness. It is an open-source Python library consisting of ready-made functions for over eighty adversarial attacks and defense methods. As technology evolves, continuous updates to such tools are necessary to address emerging challenges, such as new AI capabilities or unforeseen risks.

Standardized metrics enable easy quantification and comparison of assessment results. For example, metrics such as the Fairness Score [Agarwal et al., 2023] provide quantifiable measures of how well an AI system performs for AI fairness.

This layer facilitates in actual evaluation of the relevant aspects of AI systems based on the requirements and procedures articulated in the previous layers. It also ensures that the assessment results are comparable and universally acceptable.

### 3.5 Layer-5: Certification ecosystem

Layer-5 establishes the certification ecosystem for validating AI systems. This layer defines the processes for certifying that AI systems meet the requirements specified in the previous layers. Certification assures that AI systems comply with legal, regulatory, and ethical requirements. A certified AI system instills confidence in the deployers and users about its compliance status. Developers and auditors are the primary stakeholders in this ecosystem, working to create credible and transparent certification processes.

The certification ecosystem includes various mechanisms such as self-certification, third-party certification, and continuous monitoring. Self-certification allows organizations to assess and validate their own AI systems, while third-party certification involves independent bodies verifying compliance. Continuous monitoring ensures that AI systems remain compliant over time, adapting to changes in technology and regulations.

Several certification schemes are emerging to operationalize this layer. While some programs, like CertX, offer direct certification of AI applications against established standards [CertX, 2023], others, such as IEEE CertifAIEd™ and TÜV SÜD's AI Quality Certification Program, focus on certifying individuals in the management and quality assurance of AI systems, with the underlying goal of improving application quality [IEEE Standards Association, 2022b, TÜV SÜD, 2023]. These examples illustrate how Layer-5 supports a credible and evolving ecosystem for trustworthy AI certification, spanning both individual competencies and system-level validation.

### 3.6 Integration and Interaction of Layers

Each layer of the framework is interconnected, building upon and complementing the previous one. Layer-1's broad regulatory principles provide the foundation for the detailed requirements for each principle in the form of specific standards in Layer-2. Layer-3's standardized assessment procedures provide test methods for each requirement to operationalize these requirements, while Layer-4's tools and metrics facilitate the assessment process. Finally, Layer-5's certification ecosystem ensures that AI systems comply with the entire governance framework. This integration ensures that AI systems are rigorously evaluated, certified, and continuously monitored, contributing to their trustworthiness and reliability in real-world applications.

This layered approach ensures that AI governance is both comprehensive and adaptable. By breaking down the governance process into distinct layers, the framework allows for targeted improvements and updates, addressing specific aspects of AI governance while maintaining overall coherence. It provides a clear pathway for organizations to navigate the complexities of AI regulation and implementation, offering practical guidance at each stage of the governance process.

To validate the practical relevance of the five-layer framework, two case studies—AI fairness and AI incident reporting—are examined, each representing a different stage of maturity in AI governance.

## 4 Case Study-1: AI Fairness in Global and Indian Contexts

The first case study focuses on AI fairness, a domain that has seen significant global regulatory and academic attention. Ensuring AI fairness is crucial, as it addresses how AI systems may perpetuate or mitigate biases that affect different demographic groups. For each of the five proposed layers, the case study first examines global developments related to AI fairness. It then looks at the unique actions recently taken by India in this area to get a jurisdiction-specific view.

## 4.1   Layer-1: Laws, regulations, and policies

In the context of AI fairness, layer-1 establishes the high-level regulatory principles and legal requirements that guide the assessment of fairness in AI systems. The case study refers to the provisions of the EU AI Act on AI fairness to analyze the developments at the global level. The EU AI Act tackles AI fairness by requiring high-risk AI systems to undergo conformity assessments, including evaluation for bias and discrimination. Providers must implement risk management systems to identify, analyze, and mitigate potential biases throughout the AI lifecycle. Data used to train AI must be representative. Specific requirements apply to sensitive areas like recruitment and law enforcement, prohibiting AI that discriminates based on protected characteristics. The Act also promotes transparency and explainability to enable scrutiny of AI decision-making processes.

In addition, various multilateral organizations have released guidelines or principles for trustworthy, responsible, or ethical AI, which invariably include AI fairness and non-discrimination. Examples include the *Recommendation on the Ethics of Artificial Intelligence* released by the United Nations Educational, Scientific and Cultural Organization [UNESCO, 2022] and the *Recommendation of the Council on Artificial Intelligence* released by the Organization for Economic Co-operation and Development (OECD) [OECD, 2024]. These documents serve as guiding principles for the countries associated with these organizations.

While the EU AI Act and the various principles/recommendations released by multilateral organizations provide a general framework, they do not specify detailed procedures or benchmarks for evaluating fairness or other similar requirements. This highlights the need for more granular guidance, which the subsequent layers of the proposed framework aim to address.

Considering the specific case of India, it is noted that while India does not have any AI-specific laws, it has a National Strategy for Artificial Intelligence aiming to position India as a leader in AI while ensuring ethical and inclusive deployment [NITI Aayog, 2018]. Its approach documents, released in two parts, elaborate on the principles for Responsible AI (RAI), emphasizing fairness, accountability, and transparency (FAT) to build public trust. These documents propose highlevel principles to address AI bias, promote unbiased algorithms, and establish ethical standards, ensuring AI benefits all sections of society and aligns with the Indian values [NITI Aayog, 2021].

## 4.2   Layer-2: Standards

Layer-2 focuses on specific frameworks and standards for evaluating fairness in AI systems. This case study examines ISO/IEC 42001 [International Organization for Standardization, 2023c], ISO/IEC TR 24027:2021 [International Organization for Standardization, 2021], and the TEC Standard for Fairness Assessment and Rating of Artificial Intelligence Systems (TSFARAIS).

The ISO/IEC 42001 standard [International Organization for Standardization, 2023c] specifies requirements for an AI management system, focusing on governance, risk management, and quality assurance. While it doesn't directly address AI fairness, it establishes a framework that enables organizations to implement processes for addressing bias and promoting responsible AI development and deployment through principles like transparency and accountability.

ISO/IEC TR 24027:2021 [International Organization for Standardization, 2021] is a Technical Report adopted as a European standardization document, addressing bias in AI systems, particularly in AI-aided decision-making. It outlines techniques for measuring and assessing bias across all AI lifecycle phases, including data collection, training, continual learning, design, testing, evaluation, and use.

The TSFARAIS is a specialized framework for evaluating and rating AI systems for fairness using a risk-based, three-step approach. First, it assesses bias risk through a self-assessment questionnaire that covers risks from AI components, data types, and models. Second, it establishes thresholds to identify unfair bias, helping developers set benchmarks for testing. Third, it outlines techniques for bias testing, including process reviews, metric analysis, and scenario exploration. While the TSFARAIS provides a standardized approach for assessing AI fairness, it is currently limited to tabular data and supervised learning systems. More similar standards are needed to cover the full spectrum of data modalities and machine learning algorithms.

In India, the TSFARAIS is considered a benchmark for developing detailed standards for other aspects of AI. TEC is also developing a new standard for assessing and rating the robustness of AI systems in telecom networks and digital infrastructure along similar lines [Telecommunication Engineering Centre, 2024]. Furthermore, efforts are underway to extend the scope of the TSFARAIS to make it more comprehensive by including other data modalities such as images, unstructured text, and natural language processing (NLP), as well as extending it to large language models (LLMs) [Agarwal et al., 2025].

## 4.3 Layer-3: Standardized assessment procedures

In this case study, for layer-3, standardized procedures for conducting fairness assessments and operationalizing the frameworks and standards from layer-2 are explored. Multiple global standards bodies are developing standards focused on standardizing assessment procedures and metrics for various aspects of AI.

IEEE P3198 [IEEE Standards Association, 2022c] is developing a standard for evaluating machine learning fairness, categorizing causes of unfairness, presenting widely recognized definitions, specifying corresponding metrics, and providing test cases for evaluation procedures.

The International Telecommunication Union (ITU) Study Group 11 (SG11) is developing a draft recommendation on a framework for the quality evaluation of conversational AI (C-AI) systems [ITU-T Study Group 11, 2025]. It provides guidance on validating a voice chatbot using AI/ML key performance indicators (KPIs), including AI fairness, and lists the Fairness Score and Bias Index [Agarwal et al., 2023] as metrics for measuring AI fairness. Additionally, various ITU-T Study Groups have over 15 active work items focused on developing draft standards for assessment criteria or evaluation methods for AI-related topics.

The TSFARAIS discussed in the previous sub-section offers detailed procedures for assessing AI fairness. It includes a comprehensive checklist for classifying bias risks, step-by-step methods for evaluating data and model fairness, and scenario testing. Additionally, it provides a template for fairness evaluation outcome reports. These procedures and templates enable structured and standardized assessments, potentially facilitating fairness certifications.

Such domain-specific assessment procedures/evaluation criteria help standardize the assessment of specific aspects, such as fairness, in AI systems and are useful in ensuring consistency and comparability in fairness evaluations.

## 4.4 Layer-4: Standardized assessment tools and metrics

In this case study, layer-4 comprises the various tools and metrics for conducting AI fairness assessments.

Several open-source toolkits are widely available for assessing AI fairness, including IBM's AI Fairness 360, Microsoft's Fairlearn, Google's What-if, and Aequitas. These toolkits provide various functionalities, such as testing AI models against established fairness metrics, experimenting with different decision thresholds, and offering explanations for model predictions (like feature importance). They also enable visualizations of fairness and accuracy metrics and support the integration of fairness considerations into the model-building process. This includes techniques like debiasing and applying fairness constraints during the training phase to ensure more equitable outcomes.

Commonly used fairness metrics, such as statistical parity (demographic parity), disparate impact ratios, equal opportunity, and equalized odds, facilitate efficient and precise evaluation of AI systems. The Fairness Score [Agarwal et al., 2023], as detailed in the TSFARAIS, aggregates multiple fairness metrics into a single score, providing a comprehensive measure of an AI system's biases and quantifying its overall fairness.

In India, the Ministry of Electronics and Information Technology (MeitY) is funding the development of a tool to assess the fairness of AI models based on the TSFARAIS [IndiaAI, 2024]. This tool, named Nishpaksh, is being developed by academia in association with the standards organization TEC. Considering that developing such tools may not be financially viable, as many open-source tools are already available, the funding by the Indian Government provides a significant incentive. The project aims to either develop a new AI assessment tool from scratch or customize existing open-source toolboxes to adapt to the requirements of the TSFARAIS. One key feature of the tool being developed in India is the integration of the combined fairness metrics Fairness Score and Bias Index, which are currently absent in the open-source AI fairness tools available.

The case study highlights how these tools and metrics facilitate the assessment process. Applying standardized tools and metrics ensures that evaluation is consistent and comparable, providing reliable results that inform further improvements.

## 4.5 Layer-5: Certification ecosystem

For layer-5, this case study explores the certification ecosystem for AI fairness. Certification is crucial in ensuring compliance with fairness standards and building trust in AI systems. *SelfCertification:* Various organizations developing AI systems conduct internal assessments using the standardized procedures, tools, and metrics from Layers-3 and 4. These internal assessments provide an initial validation of fairness; however, organizations generally do not issue self-certifications based on these evaluations. Further, without external oversight, self-certification alone is insufficient to ensure accountability and regulatory compliance.

*Third-Party Certification:* Independent third-party certification programs assess AI systems for compliance with fairness and ethical principles. One of the few initiatives in this domain is IEEE CertifAIEd [IEEE Standards Association, 2022b], which evaluates AI systems based on the ethical criteria for Autonomous and Intelligent Systems (AIS). However, globally recognized AI fairness certification frameworks remain limited, leaving a gap in standardized, enforceable certification mechanisms.

In the Indian context, a review of various sources reveals that discussions are underway to promote AI fairness certification based on the TSFARAIS and the under-development AI fairness tool, Nishpaksh, in India. Additionally, there are proposals to enable leading academic institutions to establish AI fairness assessment labs for the voluntary evaluation of AI systems. These institutions are being encouraged as potential independent third-party auditors, as they are less likely to have corporate conflicts of interest. However, such initiatives are at an early stage and may require policy support to develop into a formal certification ecosystem.

The case study identifies the gap in the certification ecosystem to assure compliance with AI fairness standards/requirements. While developers often perform self-assessments, these lack independent validation. Third-party certification options remain limited and do not specifically focus on AI fairness in diverse application areas.

## 4.6 Conclusion of the Case Study-1

TABLE 2 summarizes the application of the five-layer model to a specific domain (AI fairness) while considering both global and India-specific aspects.

Table 2: The five-layer framework applied to AI fairness (Source: Authors' own work)

| Layer | Title | Global | India |
|---|---|---|---|
| 1 | Laws, regulations, and policies | The EU AI Act, OECD RAI Principles | National Strategy for Artificial Intelligence |
| 2 | Standards | ISO/IEC TR 24027:2021 | TSFARAIS** |
| 3 | Standardized assessment procedures | IEEE P3198*, ITU-T SG11 E.AIQ* | TSFARAIS** |
| 4a | Standardized assessment tools | AI Fairness 360, What-if, Fairlearn, Aequitas | Nishpaksh* |
| 4b | Standardized metrics | Demographic Parity, Equal Opportunity | Fairness Score, Bias Index |
| 5 | Certification ecosystem | IEEE CertifAIEd; self-certification | evolving |

\* : under development
\*\* : TEC Standard for Fairness Assessment and Rating of AI Systems

The case study demonstrates how the proposed five-layer framework provides a structured approach from regulatory foundations to certification to ensure fairness in AI systems. Each layer plays a distinct role in the process and contributes to a comprehensive evaluation to ensure consistency and rigor. While global standards and assessment procedures provide a foundation, the inclusion of India-specific initiatives, such as developing the AI fairness tool Nishpaksh and ongoing discussions about independent assessment labs, in the case study illustrates the framework's flexibility and relevance in regional contexts. It also highlights the importance of a dual approach, where localized efforts complement international best practices to address region-specific challenges in AI fairness.

# 5   Case Study-2: AI Incident Reporting and the Five-Layer Framework

This second case study applies the framework to a less mature domain: AI incident reporting. AI incidents can arise from unintended model failures, unexpected outcomes, biased decisionmaking, or vulnerabilities in AI-driven systems. These incidents may cause harm or pose risks to individuals and society, leading to discrimination, rights violations, cyber-attacks, or disruptions in critical infrastructure such as power grids and telecom networks [Agarwal and Nene, 2024a]. Systematic cataloging of AI incidents is essential for identifying risks, improving accountability, and strengthening governance mechanisms. Such incident data can also support research and model improvements so as to reduce the reoccurrence of similar incidents in the future [McGregor, 2021]. While efforts such as AI incident databases and sector-specific guidelines exist, there is no standardized approach to reporting and addressing AI incidents across industries. Existing initiatives, such as the *AI Incident Database* (AIID) [McGregor et al., 2021] and the *AI, Algorithmic and Automation Incidents and Controversies* (AIAAIC) [AIAAIC, 2024], operate on a voluntary basis, using different definitions and reporting mechanisms. Prior research has identified key gaps in AI incident reporting, including the lack of standard definitions, reporting formats, assessment procedures, and incentives for disclosure [Turri and Dzombak, 2023, Avinash and Manisha, 2024].

Applying the five-layer framework to AI incident reporting provides a structured approach to identifying and addressing these gaps.

## 5.1   Layer-1: Laws, regulations, and policies

Currently, AI incident reporting is not covered comprehensively under various regulations across jurisdictions. While the EU AI Act introduces reporting obligations for high-risk AI systems, it does not define a universal reporting structure. Other jurisdictions primarily address AI failures under broader cybersecurity or product liability laws, which may not capture all AI-specific risks, such as bias or emergent failures. Organizations may internally track AI failures, but without regulatory mandates or reporting incentives, there is little motivation to share incidents with outsiders. At least in critical sectors such as telecom, finance, and healthcare, a clear regulatory mandate for AI incident reporting is needed to ensure transparency and accountability in AI deployment [Agarwal and Nene, 2024a].

## 5.2   Layer-2: Standards

Presently, no globally accepted standards are there to define AI incidents or specify how they should be reported and recorded. AIID and AIAAIC each use their own definitions, so comparing incidents across different domains becomes difficult. This is unlike cybersecurity, which has standardized taxonomies for threats (e.g., CVEs for vulnerabilities). Further, aggregating incidents from different repositories is challenging because AI incident repositories lack a standardized schema [Agarwal and Nene, 2024b]. Establishing standards for defining, categorizing, and reporting AI incidents through standards organizations would improve consistency and interoperability across reporting systems.

## 5.3   Layer-3: Standardized assessment procedures

Different AI incident repositories apply their own procedures for verifying and classifying incidents [Avinash and Manisha, 2024]. AIID, for example, relies on manual curation based on media reports, while AIAAIC focuses on algorithmic impact assessment. There is no standardized methodology for assessing incident severity, root causes, or systemic risks. Developing structured assessment criteria, such as classification thresholds based on harm severity, would enable more consistent evaluation and comparison of AI incidents across sectors. This layer could also include standardized procedures for anonymizing data before sharing the reported incidents for research or other purposes.

## 5.4    Layer-4: Standardized assessment tools and metrics

Unlike cybersecurity, where automated tools can detect and log vulnerabilities in real time, AI incident reporting lacks structured assessment tools. AIID and AIAAIC rely on manual data entry, limiting scalability and real-time monitoring. Standardized tools are required to enable automated incident detection and structured reporting. Tools are also required to ensure consistency in complying with the standardized definition of the reported incidents and their correct classifications following the standardized taxonomy. Additionally, incident impact metrics would provide quantifiable insights into AI system risks.

## 5.5    Layer-5: Certification ecosystem

There is no certification system to ensure that organizations follow standardized AI incident reporting practices. Voluntary efforts like AIID and AIAAIC collect incident reports, but there is no independent validation of accuracy, completeness, or adherence to a common framework. A certification process could help address this gap by setting clear criteria for AI incident reporting. Independent auditors or regulatory bodies could certify organizations based on standardized reporting procedures, verification methods, and defined taxonomies. This would bring consistency across industries and improve trust in reported incidents. Certification could also be linked to regulatory compliance, where organizations meeting reporting standards gain recognition or incentives, reinforcing the role of structured governance in AI incident management.

## 5.6    Conclusion of the Case Study-2

Applying the five-layer framework to AI incident reporting reveals key governance gaps, from the lack of legal mandates to the absence of standardized assessment methods and reporting tools. Unlike cybersecurity, which follows established regulations and structured reporting, AI incident reporting remains voluntary and inconsistent. A structured approach, starting with legal requirements and moving through standardized definitions, assessment procedures, and compliance mechanisms, would enhance transparency, accountability, and risk management in AI governance.

# 6    Discussion and Future Work

The preceding case studies illustrate the framework's applicability across diverse AI domains. This section reflects on key insights, implementation challenges, and directions for future work.

## 6.1    Validation of the Five-Layer Framework

The selection of AI fairness and AI incident reporting as case studies was deliberate, representing one mature and one emerging domain in AI governance. This contrast enables a comprehensive validation of the framework's ability to map existing practices and uncover governance gaps.

The case study on AI fairness verifies the five-layer framework's applicability in addressing real-world AI governance challenges. Applying the framework to fairness assessment confirmed that each layer contributes to a structured, actionable approach. The progression of the framework from broad regulatory principles to detailed standards, assessment methodologies, and certification mechanisms provides a clear pathway for ensuring AI fairness while highlighting gaps, such as the absence of certification mechanisms for AI fairness assessments.

The inclusion of a country-specific scenario demonstrates the framework's adaptability to regional requirements and its ability to align global standards with local initiatives. For instance, while global fairness metrics provide a foundational assessment, the development of Nishpaksh, an AI fairness tool tailored to Indian contexts, illustrates how region-specific implementations can complement global efforts. Similarly, the absence of independent fairness assessment labs was identified as a gap under the framework's structured approach.

The case study on AI incident reporting further validates the framework by identifying governance gaps across all five layers. Unlike fairness assessment, AI incident reporting lacks legal mandates (Layer 1), standardized taxonomies and reporting mechanisms (Layer 2), structured assessment procedures (Layer 3), and scalable assessment tools (Layer 4). The absence of a certification ecosystem (Layer 5) further diminishes compliance and reporting incentives, highlighting the need for regulatory intervention, harmonized reporting standards, and independent verification mechanisms.

Existing voluntary databases such as AIID and AIAAIC illustrate these gaps. They maintain AI incident repositories without standardized definitions, verification procedures, or regulatory oversight, and focus mainly on personal incidents with limited coverage of critical infrastructure failures. The five-layer framework addresses these shortcomings by advocating legal mandates, standardized taxonomies, systematic assessment methods, and a certification mechanism, reinforcing its applicability in diagnosing governance challenges across AI domains.

Together, these case studies confirm that the five-layer framework effectively identifies governance gaps and offers a structured pathway from regulation to implementation, making it a practical tool for policymakers, standards bodies, and industry stakeholders.

## 6.2  Addressing Gaps in Implementation

The five-layer framework offers a structured approach to AI governance, but translating it into practice presents significant challenges. Many organizations, especially small and medium enterprises (SMEs), may lack the technical expertise, financial resources, or institutional support needed to implement detailed standards, assessment procedures, and certification requirements. For such entities, the compliance burden can be disproportionately high, creating barriers to participation and increasing the risk of uneven adoption across sectors.

Bridging this gap requires targeted support mechanisms. These may include subsidized certification programs, standardized assessment tools tailored to resource-constrained settings, and accessible training initiatives. Public funding or incentive schemes can further help offset compliance costs, particularly for startups and SMEs operating in regulated AI domains.

A related challenge is ensuring consistent and widespread adoption across jurisdictions. The framework's success depends on the active engagement of governments, regulatory bodies, industry stakeholders, and civil society in integrating its components into existing governance ecosystems. However, weak enforcement mechanisms, inconsistent standards, and limited incentives can hinder implementation. Addressing this requires regulatory harmonization, alignment with international norms, incentives for voluntary adoption, and the establishment of shared infrastructures, such as independent audit bodies and regional certification hubs.

Future work should focus on developing implementation strategies that enhance the framework's practicality and scalability. This includes creating support systems for organizations with limited capacity, promoting cross-sector knowledge exchange, and embedding the framework within existing regulatory and market structures to reduce friction and encourage broader uptake.

## 6.3  Need for Coordinated Oversight Across Layers

While the five-layer framework emphasizes distributed ownership across different governance functions, effective coordination among layers remains essential. However, establishing a single centralized regulator to oversee all five layers across domains may be impractical and counterproductive. As the framework progresses from high-level legal principles to domain-specific tools and certification mechanisms, governance responsibilities naturally diverge. For instance, while AI fairness principles may be consistent at the policy level, assessment procedures or certification requirements would differ significantly between sectors like healthcare and telecommunications, each typically overseen by their respective regulators.

The framework, therefore, supports a multi-actor governance model, where governments, standards bodies, domain regulators, academia, and industry stakeholders each play distinct and complementary roles as indicated in Table 1. Encouraging collaboration, interoperability, and shared infrastructure rather than consolidation under a single authority is better aligned with the complexity and diversity of AI applications across sectors.

## 6.4  The Role of Ethical and Societal Considerations

The five-layer framework must continuously incorporate ethical and societal considerations in AI governance. As AI systems become widespread, their potential to influence social structures, economic opportunities, and human rights grows. Addressing these concerns requires an evolving framework that takes care of ethical safeguards at every stage.

Future research should focus on embedding ethical principles within each layer. For example, future standards and assessment procedures should evaluate both technical fairness and broader societal impacts. Certification mechanisms could include ethical audits to assess an AI system's alignment with human values and social norms. Strengthening these aspects will help ensure AI systems are not just technically robust but also socially responsible.

## 6.5 International Cooperation and Harmonization

AI governance is a global challenge since AI systems developed in one jurisdiction often operate across multiple borders. Current regulatory approaches like the EU AI Act are region-specific, and policies worldwide do not align with each other. This fragmentation creates inconsistencies, making it difficult for organizations that operate globally to comply with diverse regulations and for regulators to enforce standards consistently. Greater international cooperation and harmonization of governance standards would significantly enhance the framework's effectiveness.

International cooperation could facilitate the development of global standards for AI governance. Standards bodies such as ITU and ISO can help align regulatory expectations and certification processes. Standardized approaches would reduce compliance burdens while improving oversight and enforcement. Collaboration among countries can promote the exchange of best practices and the creation of shared resources, such as assessment tools and certification frameworks, that support AI governance across regions.

The discussion reveals both strengths and challenges in the five-layer framework for AI governance. Although it comprehensively connects regulatory principles with practical implementation and certification, notable challenges remain. Future work should expand the framework to cover additional governance aspects, develop standards for emerging technologies, improve the certification ecosystem, and strengthen international cooperation. Integrating ethical and societal considerations at every layer is essential to govern AI systems in ways that promote trust, accountability, and positive social outcomes.

## 7 Conclusion

This paper presents a five-layer structured framework for AI governance, addressing the need for a structured and systematic approach to regulating and managing AI systems. The framework spans from high-level regulatory frameworks, such as the EU AI Act, to specific tools and certification mechanisms, providing a clear and actionable pathway for organizations to follow. The framework ensures that AI governance is addressed comprehensively from principles to practice by progressively narrowing the focus from broad principles to specialized procedures.

The AI fairness case study highlights the adaptability of the framework in addressing both global fairness measures and region-specific initiatives, such as fairness certification in India. It reinforces the need for structured assessment methods and certification mechanisms to bridge the gap between evaluation and enforcement. The AI incident reporting case study demonstrates the framework's relevance in structuring governance for AI incidents, emphasizing the need for legal mandates, standardized reporting procedures, and certification mechanisms to improve transparency and accountability. Together, these case studies validate the five-layer framework as a comprehensive approach to AI governance, demonstrating its adaptability across regulatory and operational contexts and its effectiveness in translating regulatory principles into practice.

Robust governance mechanisms become essential as AI systems evolve and integrate into various aspects of society. The five-layer framework provides a roadmap for policymakers, regulators, and industry stakeholders to develop and implement more effective AI governance frameworks. It emphasizes the importance of moving beyond high-level regulations to include detailed standards, assessment procedures, and certification processes, ensuring that AI systems not only comply with ethical and legal standards but also operate in a responsible manner.

In conclusion, this paper contributes significantly to the field of AI governance by proposing a framework that bridges the gap between regulatory aspirations and practical implementation. By integrating high-level regulations with detailed procedural guidance, the five-layer framework offers a holistic solution to the AI governance challenges in an increasingly complex and dynamic technological landscape. Its structured approach can aid in developing more accountable and transparent AI systems, building trust and reliability in their deployment across all sectors.